\documentclass[aps,twocolumn,showpacs]{revtex4}
\usepackage{graphicx}
\usepackage[all]{xy}
\usepackage{amsmath}
\usepackage{amssymb}
\newcommand{\be}{\begin{equation}}
\newcommand{\ee}{\end{equation}}
\newcommand{\ben}{\begin{eqnarray}}
\newcommand{\een}{\end{eqnarray}}
\newcommand{\bes}{\begin{subequations}}
\newcommand{\ees}{\end{subequations}}

\newcommand{\ov}{\overline}
\newcommand{\bb}{\bibitem}

\newcommand{\LL}{{\cal L}}
\newcommand{\LX}{{\cal L}_X}

\newcommand{\LXij}{{\cal L}_{X_{ij}}}
\newcommand{\LP}{{\cal L}_{\phi}}
\newcommand{\LPi}{{\cal L}_{\phi_i}}

\begin{document}

\title{First-order framework and generalized global defect solutions}
\author{D. Bazeia, L. Losano, and R. Menezes}
\affiliation{Departamento de F\'{\i}sica, Universidade Federal
da Para\'{\i}ba, 58051-970 Jo\~ao Pessoa, PB, Brazil}
\date{\today}
\begin{abstract}
This work deals with defect structures in models described by scalar fields. The investigations focus on generalized models,
with the kinetic term modified to allow for a diversity of possibilities. We develop a new framework, in which we search for first-order
differential equations which solve the equations of motion. The main issue concerns the introduction of a new function, which works like
the superpotential usually considered in the standard situation. We investigate the problem in the general case, with an arbitrary number of fields,
and we present several explicit examples in the case of a single real scalar field.
\end{abstract}
\maketitle

{\bf I. Introduction.} The first recent observations collected in \cite{results} have led us with the intriguing fact that the Universe is presently undertaking accelerated expansion. These informations directly contributed to establish some important advances in Cosmology, one of them being the presence of dark energy, which is introduced to respond for more then 2/3 of the contents of the Universe -- see, e.g., Refs.~{\cite{reviews}} for some recent reviews of the subject.

The presence of dark energy has opened some distinct routes of investigations. The general idea starts with the recognization that the standard cosmological model is the $\Lambda CDM$ model, which describes gravity with the standard scalar curvature $R,$ with the inclusion of the cosmological constant and non relativistic dust-like matter. The dark energy modification of this scenario may be considered with the inclusion of several distinct possibilities, the two most active directions of investigations being the quintessence way, in which one considers the possibility of changing the standard scenario with the inclusion of scalar fields, and the $f(R)$ way, in which one modifies gravity, changing the contribution of the scalar curvature $R$ to $f(R)$, with $f(R)$ being a non trivial function of the scalar curvature.

In the present work we shall follow the quintessence way, that is, we shall deal with standard gravity but changing the scalar field dynamics. The use of non canonical models has appeared both in early time inflation and in dark energy scenarios to describe the current phase of the Universe \cite{adm,chiba,ams}.
But here, however, we shall make an important reduction which concerns the absence of geometrical degrees of freedom. That is, we shall deal with flat space-time.
The idea of freezing out the geometrical degrees of freedom is strategical: we shall first deal with general scalar field dynamics, and then, in another work, we shall pursue the yet more general scenario, activating the geometrical degrees of freedom \cite{bglm}. The final goal is to investigate the general model, in which we mix quintessence and $f(R),$ but here we first deal with the simpler case, where the scalar fields are allowed to interact and self-interact in the flat space-time. 

Such investigations are directly related to the recent investigations \cite{bab,blmo,adam,bab2,Adam,O} where one discusses general properties of topological defects such as domain wall, strings and monopoles \cite{books}. These investigations have dealt with formal aspects of unidimensional topological solutions \cite{bab} and specific study on kinks and the corresponding linear stability \cite{blmo}, including the presence of compactons for a specific class of models \cite{adam}. More recently the case of local vortex-like defects have been investigated in \cite{bab2} and similar modifications with higher-order scalar kinetic terms in the braneworld scenario with a single extra dimension \cite{Adam,O}.

Although the presence of new non linear effects which come from the generalized dynamics complicates the search for analytical solutions, we take motivation from the previous investigations \cite{fo} to consider the possibility of finding first-order differential equations that solve the corresponding equations of motion.
This possibility is implemented with the introduction of a general function, $W,$ which in principle depends on all the scalar fields and nicely leads to the presence of first-order equations. As usual, the first-order equations simplify the investigation as we show below, where we split the subject in two parts,
one dealing mainly with the formal aspects of defects and the other, which is dedicated to the study of specific examples.
 
{\bf II. Generalities.} We start with a model given by a non standard action, described by the set of $n$ scalar fields $\{\phi_1,\phi_2,...,\phi_n\}$ in the two-dimensional space-time
\be
\label{action1}
S=\int d^2x\;  {\cal L}(\phi_i, X_{jk}) 
\ee
where $i,j,k=1,2,...,n$. We are considering the metric $(+-),$ and the scalar fields are all dimensionless, with dimensionless space and time coordinates, and coupling constants. Here the quantities $X_{ij}$ are defined by 
\be
X_{ij}=\frac12 \,\partial^\mu\phi_i \partial_\mu\phi_j
\ee

This is the action which controls the generalized models. It allows obtaining the energy-momentum tensor in the form 
\be
T_{\mu\nu}=\LXij \partial_\mu \phi_i \partial_\nu \phi_j\,  - g_{\mu\nu} \LL
\ee
and we are using the notation: ${\cal L}_a=\partial{\cal L}/\partial a,$ that is, $\LXij=\partial \LL/\partial X_{ij}$ and $\LPi=\partial \LL / \partial \phi_i$, etc. We shall be searching for static solutions, and for static fields we can write the energy and stress densities, $\rho(x)=T_{00}$ and $\tau(x)=T_{11},$ as respectively
\bes\ben
\rho(x)&=&-\LL
\\ \label{stress23}
\tau(x)&=&{\cal L}_{X_{ij}}\phi_i^\prime\phi_j^\prime+{\cal L}
\een\ees

The equations of motion for the $n$ real scalar fields $\phi_i=\phi_i(x,t)$ have the non standard form
\be\label{eqmot}
\partial_\mu \left(\LXij \partial^\mu \phi_j \right)={\cal L}_{\phi_i}
\ee
We expand these equations to get
\be
G^{\alpha\beta}_{ij}\partial_\alpha \partial_\beta \phi_j + 2 X_{jl} {\cal L}_{X_{ij}\phi_l} - {\cal L}_{\phi_i} = 0
\ee
where
\be
G^{\alpha\beta}_{ij}={\cal L}_{X_{ij}} \eta^{\alpha\beta} + {\cal L}_{X_{il}X_{jm}} \partial^\alpha \phi_l \partial^\beta \phi_m
\ee
For static solutions $\phi_i=\phi_i(x),$ the equations of motion reduce to
\be
-\left(\LXij \phi_j^\prime \right)^{\prime} = {\cal L}_{\phi_i}
\ee
or better,
\be\label{eqmotionstatic}
({\cal L}_{X_{ij}}  +2 {\cal L}_{X_{il}X_{jm}}  X_{lm})\phi_j^{\prime\prime} = 2 X_{jl} {\cal L}_{X_{ij}\phi_l} - {\cal L}_{\phi_i} 
\ee

These equations can be integrated to give $\LL - 2 \LXij X_{ij} = C$, where $C$ is an integration constant which can be identified as the stress density \eqref{stress23}. Stability of the static solution requires the vanishing of $C,$ that is, stability requires the stressless condition \cite{blmo}. Therefore we write
\be\label{cc}
\LL - 2 \LXij X_{ij} =0	
\ee
Notice that this equation depends of the scalar fields and their first derivatives. Therefore, it is a first-order equation, but it is a constraint equation, which we name the stressless constraint.

We now deal with the search of first-order differential equations. We consider the energy density, which for the stressless solutions can be written as 
\be
\rho= - \LL= \LXij \phi^\prime_i  \phi^\prime_j
\ee
In this case, if we introduce the new function $W=W(\phi_1,\phi_2,...,\phi_n)$ such that
\be\label{W}
\LXij \phi^\prime_j = W_{\phi_i}
\ee
we obtain the energy density in the form 
\be\label{energyW}
\rho=W_{\phi_i}\phi_i^\prime=\frac{dW}{dx}
\ee
in a way such that the energy can be written as the variation of $W$, that is 
\ben
E=\Delta W&=&W(\phi_1(\infty),\phi_2(\infty),...,\phi_n(\infty))\nonumber
\\
&-&W(\phi_1(-\infty),\phi_2(-\infty),...,\phi_n(-\infty))
\een
In order to circumvent problem with unstable solutions, here we assume that $W$ is a ${\cal C}$$^1$ function.
 
We substitute (\ref{W}) in the equations of motion (\ref{eqmotionstatic}) to get 
\be\label{condstress}
W_{\phi_i\phi_j}\phi_j^{\prime}=-{\cal L}_{\phi_i}.
\ee
which is also a set of first-order differential equations which solve the equations of motion. Since the set of equations \eqref{W} is a first-order set of equation, we name the procedure the first-order framework for the scalar field models under investigation. In this sense, the above investigation falls into the program shown in \cite{fo}, in which we dealt with a diversity of models, searching for first-order differential equations which solve the corresponding equations of motion. The first-order equations are also important to investigate linear stability, because the appearance of $W$ eases the general investigation and helps to factorize the Schr\"odinger-like Hamiltonian into the two first-order differential operators $S$ and $S^\dag,$ which we explicitly construct below.
   
The above calculations generalize the standard scenario, which appears when the scalar fields evolve under usual dynamics. We note that the energy does not depend on the explicit form of the solution, but it is given in terms of $W$, calculated at the asymptotic values of the fields. The non vanishing of the energy indicates the presence of non trivial field configuration having non trivial topology which ensures existence and stability of the solution. 

The stability of the static solution can be inferred by the presence of the topological current, defined in terms of the function $W$ introduced above:
\be
j_T^\mu=\epsilon^{\mu\nu}\partial_\nu W
\ee
This definition is inspired in \cite{bazeia}, and it is constructed to make the topological charge directly related to the energy of the static solution. We see that for static solution $j_T^0=dW/dx,$ so that $Q_T$ is directly related to $\Delta W$.

We now focus on linear stability. We consider $\phi_i(x,t)=\phi_i(x)+\eta_i(x,t),$ supposing that $\eta_i(x,t)$ are small fluctuations around the static solution. In this case we get, going up to first-order in the fluctuations,
\be
X_{ij}=X_{ij}+\bar{X}_{ij}
\ee 
with
\be
\bar{X}_{ij}=\frac12\partial_\mu\phi_i\partial^\mu\eta_j+\frac12\partial_\mu\phi_j\partial^\mu\eta_i
\ee
The procedure gives
\bes\ben
{\cal L}_{\phi_i}&\to&{\cal L}_{\phi_i}+{\cal L}_{\phi_i\phi_j}\eta_j+{\cal L}_{\phi_iX_{jk}}\bar{X}_{jk}
\\
{\cal L}_{X_{ij}}&\to&{\cal L}_{X_{ij}}+{\cal L}_{X_{ij}\phi_k}\eta_k+{\cal L}_{X_{il}X_{mj}}\bar{X}_{lm}
\een\ees
We use these expressions into the equations of motion \eqref{eqmot} to get
\ben
&&\partial_\mu ({\cal L}_{X_{ij}} \partial^\mu \eta_j + {\cal L}_{X_{ij} X_{jm}}\partial^\mu \phi_j X_{lm})\nonumber
\\
&&\;\;\;=\left[{\cal L}_{\phi_i \phi_j} - \partial_\mu ({\cal L}_{X_{im} \phi_j}  \partial^\mu \phi_m )\right] \eta_j
\een
In the case of static solutions one gets
\ben
{\cal L}_{X_{ij}}\ddot\eta_{j}&-&\left(\left({\cal L}_{X_{ij}}+{\cal L}_{X_{il}X_{jm}}X_{lm}\right)
\eta^\prime_j\right)^\prime\nonumber\\
&=&\left({\cal L}_{\phi_i \phi_j}+({\cal L}_{X_{im} \phi_j} \phi_m)^\prime\right) \eta_j
\een

We suppose that the fluctuations are given by
\be
\eta_i(x,t)=\eta_i(x)\cos(\omega t)
\ee
in order to get
\ben
&-&\left[\left({\cal L}_{X_{ij}}+{\cal L}_{X_{il} X_{jm}} X_{lm}\right) \eta^\prime_j\right]^\prime\nonumber\\
&=&\left[{\cal L}_{\phi_i \phi_j}+({\cal L}_{X_{im} \phi_j} \phi_m)^\prime+\omega^2{\cal L}_{X_{ij}}\right]\eta_j
\een

This equation has the general form
\be\label{fluct}
(-a_{ij}\eta_j^\prime)^\prime=b_{ij}\eta_j
\ee 
We can modify the above equation into the Schr\"odinger-like equation
\be
\left(-\delta_{ij}\frac{d^2}{dz^2}+U_{ij}\right)u_j=w^2u_i
\ee
where the potential $U$ is now a matrix which depends on the matrix $S$ and $R,$ introduced as folows: in \eqref{fluct} we change $\eta_i(x)$ into $u_i(z),$ such that
\bes\label{C}
\ben
\eta_i&=&S_{ij}u_j
\\
dx&=&\frac{dz}{R}
\een\ees
In this case, the Schr\"odinger-like equation requires that
\bes\label{A}
\ben
2a_{ij} R \frac{dS_{jk}}{dz}+ \frac{d(a_{ij} R)}{dz} S_{jk}&=&0
\\
R^{-2}S^{-1}_{il}a^{-1}_{lm}{\cal L}_{X_{mj}}&=&\delta_{ij}
\een\ees
The general study is awkward. The issue here is that it is better to investigate linear stability in a case by case manner. We will return
to this below, where we examine the given examples specifically.   

The above investigations are done on general grounds. To see how it generalizes the standard situation, let us suppose that the Lagrange density has the very specific form
\be
{\cal L}=M^{ij}X_{ij}-V
\ee
where $M^{ij}=M^{ij}({\phi_i})$ and in principle $M$ may also depend on the scalar fields. In this case we have that, using the first-order equation \eqref{W}
\be
\phi_i^\prime= M^{-1}_{ij} W_{\phi_j}
\ee 
The stressless constraint \eqref{cc} leads to
\be
V=\frac12 W_{\phi_i}M^{-1}_{ij}W_{\phi_j}
\ee
which shows that the function $W$ is a generalization of the superpotential which appears in the standard scenario. To be explicit, let us consider the standard case, in which $M$ is constant, $M^{ij}=\delta^{ij}$. Here we have
\be
{\cal L}=\frac12\delta^{ij}\partial_\mu\phi_i\partial^\mu\phi_j-V(\phi)
\ee
Also, from \eqref{W} we can write
\be
\phi_i^\prime=\delta_{ij}W_{\phi_j}
\ee
and from \eqref{cc} we have to have
\be
V(\phi)=\frac12 \delta_{ij} W_{\phi_i}W_{\phi_j}
\ee
This is the standard scenario and this the reason why we use $W$ to represent the new function that we had to introduce in \eqref{W}. As we know, in the standard case we can include fermions to get to the supersymmetric extension of the model, and there $W$ is named the superpotential. This poses a nice issue, which concerns the inclusion of fermions in the case of generalized dynamics, to get to the supersymmetric model -- this is an interesting issue which is presently under consideration.

The very interesting feature of the present generalization is that we get to first-order differential equations which solves the equations of motion,
even though we do not specify the potential. This is a strong result, which shows that we can get to the first-order framework even though we do not know explicitly how the scalar fields interact formally.

{\bf III. Applications.} Let us now move toward applications. For simplicity, we consider models described by a single real scalar field, with the equation of motion being written in terms of $W.$ The general model is described by
\be
\LL=\LL(X,\phi)
\ee
This case was already investigated in \cite{bab,blmo,adam} under specific conditions, but here we will introduce other improvements. The equation of motion and the stressless conditions are given by
\be
-\left[\LX \phi^\prime\right]^\prime = \LP
\ee
and
\be
\LL - 2 \LX X = 0 
\ee
The first-order equation has the form
\be
\LX \phi^\prime= W_\phi
\ee
This is the general case. The first-order equation supports kinklike solutions, and can have the topological charge shown before. The study of linear stability
follows the general investigation given in the former section. We solve the equations \eqref{C} and \eqref{A} to get 
\be
dx=\frac{dz}{A} \,\,\,\,\,\,\,\,{\rm and} \,\,\,\,\,\,\,\, \eta=\frac{u}{\sqrt{{\cal L}_X A}}
\ee
where $A^2=(2{\cal L}_{XX} X + {\cal L}_X)/{\cal L}_X$. This allows writing the Schr\"odinger-like equation
\be
-u_{zz} + U(z) u = \omega^2 u
\ee
where
\be
U(z)=\frac{(\sqrt{A{\cal L}_X})_{zz}}{\sqrt{A{\cal L}_X}} -\frac{1}{{\cal L}_X} \left[{\cal L}_{\phi\phi}+\frac{1}{A} \left({\cal L}_{\phi X} \frac{\phi_z}{A}\right)_z\right]
\ee
See also Ref.~{\cite{blmo}} for other details.

We illustrate the general behavior with the model
\be\label{xx2}
\LL = X -	\alpha X^2 - V(\phi)
\ee
or equivalently 
\be
\LL=X+\alpha X |X|-V(\phi)
\ee
for $\alpha>0$ real and positive parameter. In this case, the equation of motion is
\be
(1 - 3\alpha \phi^{\prime2})\phi^{\prime\prime}=V_\phi
\ee
and the first-order and stressless equations are given by, respectively
\bes\label{eq30}
\ben
\phi^\prime + \alpha \phi^{\prime 3} &=& W_\phi\\
\frac12 \phi^{\prime2} + \frac{3\alpha}4 \phi^{\prime4} &=& V(\phi) 
\een
\ees
The first equation is an algebraic equation of third degree in $\phi^\prime$. The only real solution is
\be
\phi^\prime = G_\alpha(W_\phi)
\ee
where the function $G_\alpha(W_\phi)$ is such that
\be
G_\alpha(v)=\frac{g_\alpha(v)}{6\alpha} -\frac{2}{g_\alpha(v)}
\ee
and
\be
g_\alpha(v)=(54\alpha^2\,v +6\sqrt{3}(16\,\alpha^3 + 27 \alpha^4\,v^2)^{1/2})^{1/3}
\ee
We now use the stressless constraint to write the potential in the form 
\be	
V(\phi)=\frac12 G^2_\alpha(W_\phi) + \frac{3\alpha}4 G^4_\alpha(W_\phi) 
\ee

The advantage of the introduction of the function $W$ is that we know a priori the energy of the topological solutions, since their asymptotic values are found from the relation $G_\phi(W_\phi)=0.$ In this case, $V(\phi)$ is $\alpha$-dependent. The inclusion of the quartic term changes some of the features of the solutions, as the width and energy. We illustrate this with the function
\be\label{Wphi4}
W=\phi-\frac13\phi^3
\ee
We use this $W$ to plot in Fig.~1 the potential, solution and energy density 
\be
\rho(x)=G_\alpha^2(W_\phi) + {\alpha}\,G_\alpha^4(W_\phi)
\ee
which are depicted for some specific values of the parameter $\alpha$, with $\alpha=0$ being the thicker line. The first-order equation $\phi^\prime=G_\alpha(1-\alpha^2)$ is solved numerically. In Fig.~1, we see that both the potential, and the width and energy density of the
solution change with $\alpha,$ showing how the modification introduced by $\alpha$ change the specific features of the defect solutions.

\begin{figure}[ht!]
\includegraphics[width=5cm]{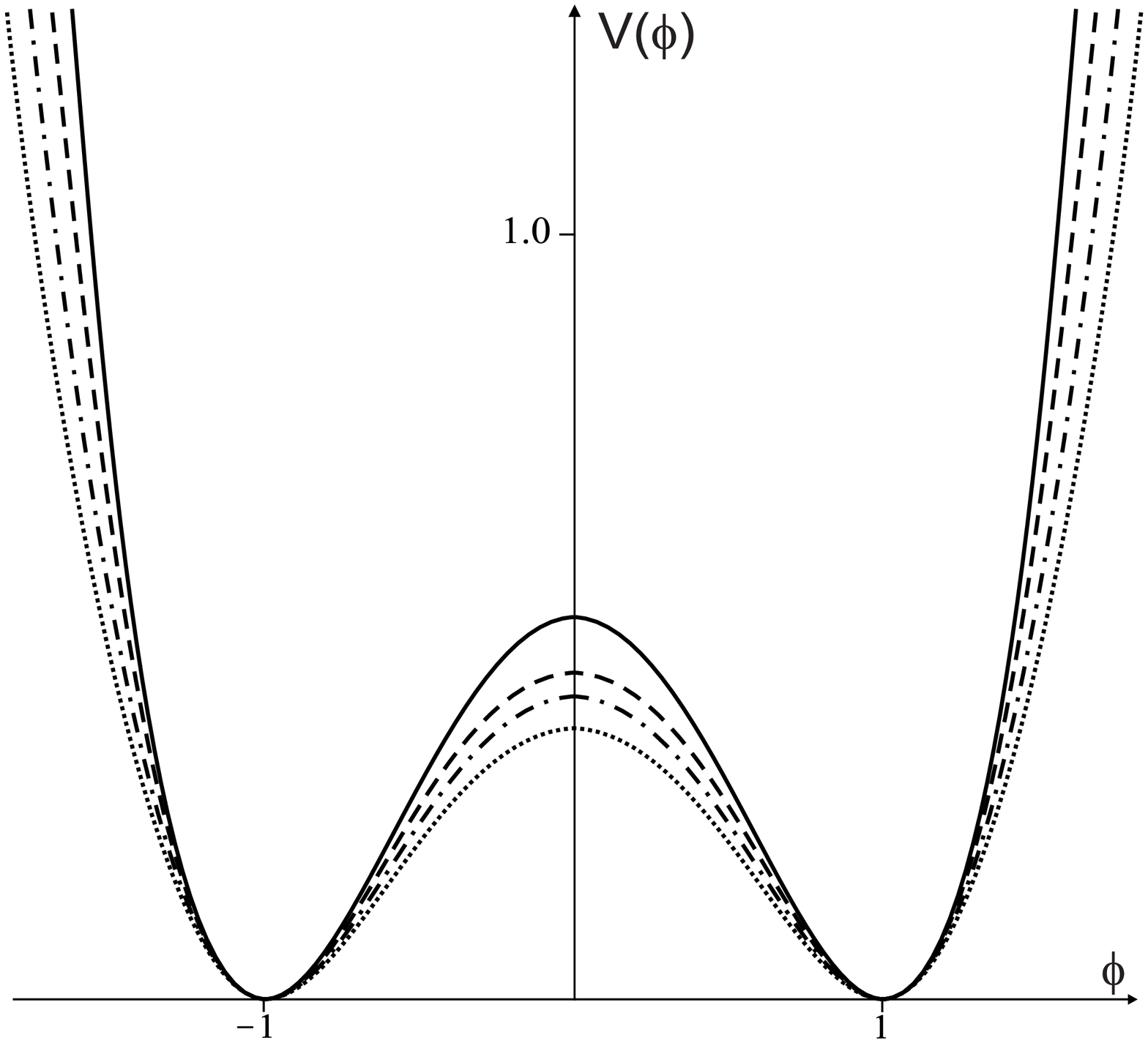}\vspace{0.3cm}
\includegraphics[width=5.2cm]{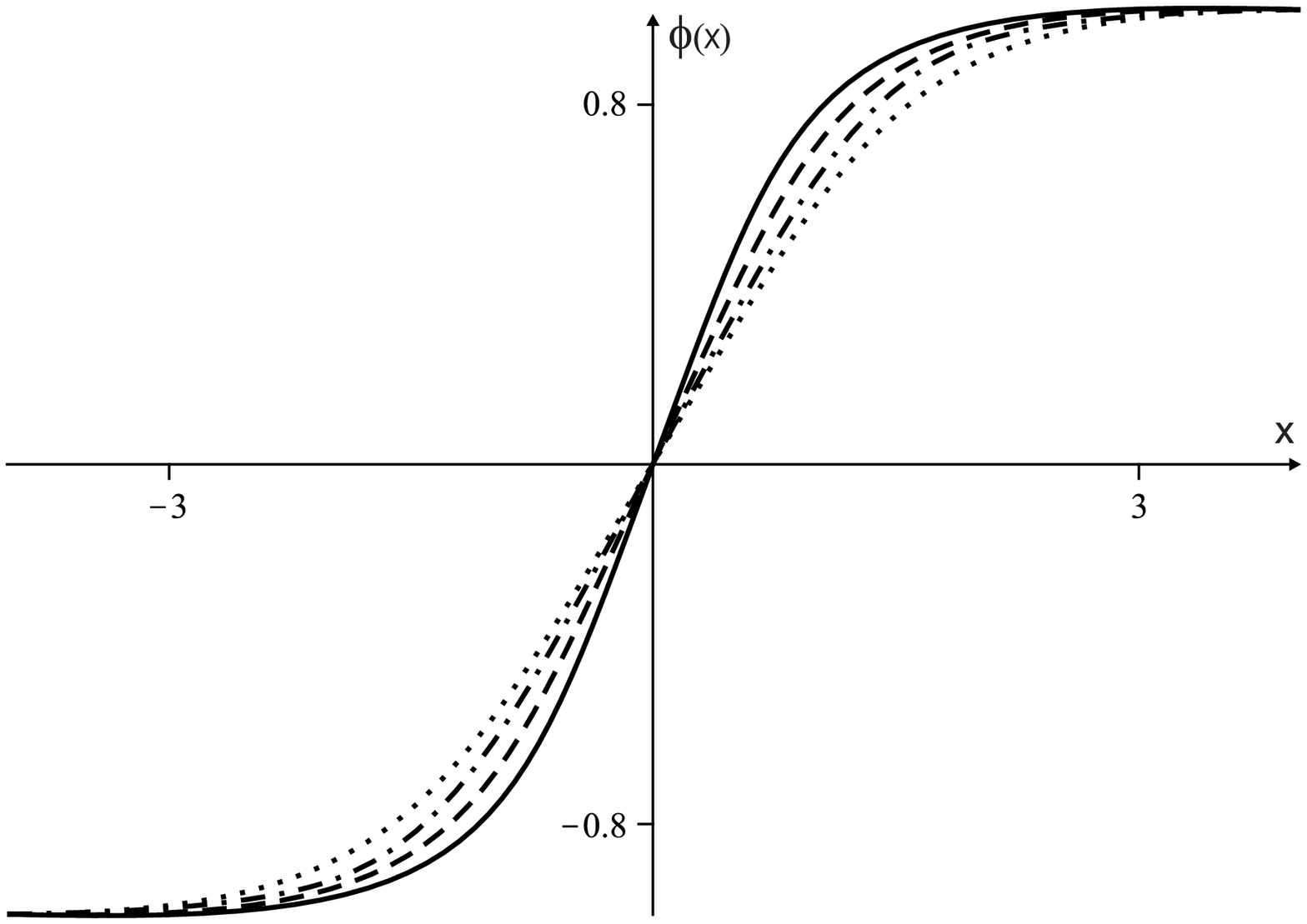}\vspace{0.3cm}
\includegraphics[width=5cm]{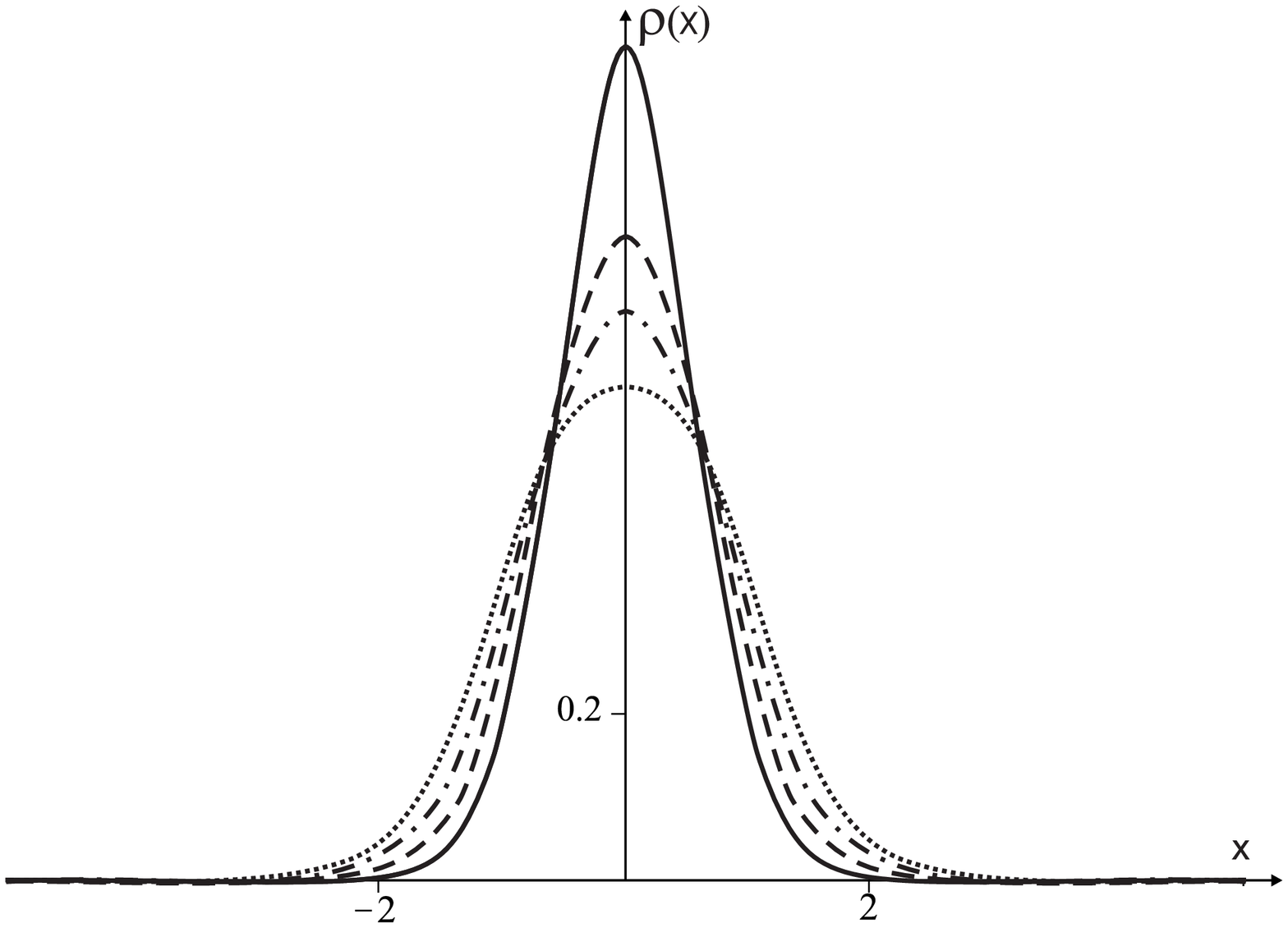}
\caption{Plots of the potential (upper panel), kinklike solution (middle panel) and energy density (lower panel) for the model with $W=\phi-\phi^3/3$, with several values of $\alpha.$ The case $\alpha=0$ is shown with the solid line, and the cases with $\alpha=0.5, 1.0, 2.0,$ appear with dashed, dot-dashed and dotted line, respectively.}
\end{figure}

The configurations have asymptotic values at $-1$ for $x\to -\infty,$ and $+1$ for $x\to \infty.$  Therefore, as the energy only depends on the asymptotic values, $E=W(\phi(+\infty))-W(\phi(-\infty))$, the solution have the same energy of the standard $\phi^4$ case, which is $E=4/3.$ In Ref.~\cite{blmo}, we have studied models in which the potential had no explicit dependence on $\alpha$, as well as the solution. Here, in the present models we see that it is the energy that does not change with $\alpha$. In this sense, the present investigation adds to the former scenario of {\cite{blmo}} another interesting possibility, in which the energy of the topological solution is fixed, although its specific form may change together with the potential.

If we consider $\alpha$ very small we can go further and write, up to first order in $\alpha$
\be
G_\alpha(v)=v-\alpha v^3 \label{Gfirst}.
\ee 
which is good approximation for $v$ limited. We note that $G_0(v)=v$ and $G_\alpha(0)=0$, and in this case the equations \eqref{eq30} lead to
\be\label{eq1a}
\phi^\prime=W_\phi-\alpha W^3_\phi
\ee
and so the potential becomes
\be
V(\phi)=\frac12 W^2_\phi-\frac14 \alpha W^4_\phi
\ee
We then integrate \eqref{eq1a} to get to
\be
\phi(x)=\phi_0(x)-\alpha W_{\phi_0} W(\phi_0)
\ee
where $\phi_0(x)$ is the solution for $\alpha\to0.$ The corresponding energy density is
\be
\rho(x)=\rho_0(x)\,\left(1-\alpha (W^2_{\phi_0}+2WW_{{\phi_0}{\phi_0}})\right)
\ee
where $\rho_0(x)=W^2_{\phi_0}$.

The explicit form $W=\phi-\phi^3/3$ leads to the standard results
\bes\ben
\phi_0(x)&=&\tanh(x)
\\
\rho_0(x)&=&{\rm sech}^4(x)
\een\ees
which are valid for $\alpha$ equal to zero; for $\alpha$ very small, we get the new analytical results up to first order in $\alpha$
\bes\ben
\phi(x)\!=\!\tanh(x)\!-\!\frac{\alpha}{3}\tanh(x){\rm sech^2(x)}\!\left(2\!+\!{\rm sech}^2(x)\!\right)\label{phi11}
\\
\rho(x)\!=\!{\rm sech}^4(x)\!\left(\!1\!-\!\frac{\alpha}{3}\left(7\,{\rm sech}^4(x)\!+\!4\,{\rm sech}^2(x)\!-\!8\right)\!\right)\label{rho11}
\een\ees
There is no contribution to the energy from the $\alpha-$dependent terms in $\rho$ in \eqref{rho11}, and this is in accordance with the fact that the energy is $E=\Delta W=4/3$ and only depends on $W$ and the asymptotic values of the field, which do not depend on $\alpha$, as we see from $\eqref{phi11}.$

We use 
\bes\ben
dx&=&\left(1+\alpha W_\phi^{2} \right)dz \\
\eta&=&(1-\alpha W_\phi^2) u
\een\ees
in order to find the potential of the Schr\"odinger-like equation 
\be\label{Pot1Sch}
U(z)=W_\phi W_{\phi\phi\phi} +W_{\phi\phi}^2
\ee
and for $W$ given by \eqref{Wphi4} we get
\ben
U(z)&=&4-6\;{\rm sech}^2(z)\nonumber
\\
&&\left(1-\frac23 \alpha\;({\rm sech}^4(z)-2\;{\rm sech}^2(z)+1)\right)
\een
and the zero mode is given by, explicitly
\ben
u_0(z)&=&\frac34{\rm sech}^2(z)\nonumber
\\
&&\left(1-\frac23 \alpha\;(5\;{\rm sech}^4(z)-\;{\rm sech}^2(z)-1)\right)
\een

We can consider another model, given by
\be\label{xmodx}
{\cal L}=X|X|-V(\phi)
\ee
or yet, we make it more general and consider the case
\be
{\cal L}=\frac{2^{n-1}}{n}X|X|^{n-1}-V(\phi)
\ee
where $n=1,2,...$ is non vanishing, positive integer. The case $n=1$ leads to the standard model, and for $n=2$ we get to \eqref{xmodx}, which is the model recently investigated in \cite{adam,Adam}.

The equation of motion for $n$ arbitrary is given by
\be
(2n-1)\phi^{\prime2(n-1)}\phi^{\prime\prime}=\frac{dV}{d\phi}
\ee
We use the first-order equation to write
\be
\phi^\prime=W_\phi^{\frac1{2n-1}}
\ee
and the stressless condition leads to the potential
\be
V(\phi)=\frac{2n-1}{2n}\,W_\phi^{\frac{2n}{2n-1}}
\ee
and we also have 
\be
\rho=\phi^{\prime2n}=W_\phi^{\frac{2n}{2n+1}}
\ee
We also have $A^2=2n-1$ and the quantity 
\be
\sqrt{{\cal L}_X A}=(2n-1)^{1/4}2^{\frac{n-2}{2}}|\phi^\prime|^{n-1}
\ee

Since we are searching for topological solutions, the scalar field is supposed to be monotonic, so we take $\phi^\prime>0.$ Also, we change from $dx\to dz$ and $\eta\to u$ in order to get
\bes\ben
dx&=&\frac1{\sqrt{2n-1}} dz
\\
\eta&=&(2n-1)^{-1/4} 2^{\frac{2-n}{2}} (\phi^{\prime})^{1-n} u
\een\ees
and so the Schr\"odinger-like equation can be written in the form
\be
Hu(z)=\left(-\frac{d^2}{dz^2}+U(z)\right)u(z)=w^2u(z)
\ee
where the potential is
\be
U(z)=-\frac{n(n-2)}{2n-1}W_\phi^{-\frac{4(n-1)}{2n-1}}W^2_{\phi\phi}+nW_\phi^{-\frac{2n-3}{2n-1}}W_{\phi\phi\phi}
\ee
We factorize this equation with the first-order operators $S$ and $S^\dag$ such that $H=S^\dag S,$ with
\be
S=-\frac{d}{dz}+\frac{n}{\sqrt{2n-1}}W_\phi^{-\frac{2(n-1)}{2n-1}} W_{\phi\phi}
\ee
In this case the normalized zero mode is given by
\be
\eta_0(z)={\sqrt{\frac{A}{E}}}\; W_\phi^{\frac{n}{2n-1}}
\ee
where $E$ is the energy of the corresponding solution.

We now consider the case $n=2$ explicitly. Here the equation of motion becomes
\be
3\phi^{\prime2}\phi^{\prime\prime}=\frac{dV}{d\phi}
\ee
The first-order equation is
\be
\phi^\prime=W_\phi^{1/3}
\ee
and the stressless condition leads to the potential
\be
V(\phi)=\frac34\,W_\phi^{4/3}
\ee
We also have $A^2=3,$ and the fluctuations about the kinklike solution with $\phi^\prime$ non negative leads to the Schr\"odinger-like equation
\be
Hu(z)=\left(-\frac{d^2}{dz^2}+2W_\phi^{-1/3}W_{\phi\phi\phi}\right)u(z)=w^2u(z)
\ee 
It can be factorized with the first-order operators $S$ and $S^\dag,$ with
\be
S=-\frac{d}{dz}+\frac{2\sqrt{3}}{3}W_\phi^{-2/3}W_{\phi\phi}
\ee
This result shows that there is no bound state with negative eigenvalue, and the normalized zero mode is given by
\be
u_0(z)={\sqrt{\frac{A}{E}}}\;W_\phi^{2/3}
\ee

We can further illustrate the above case with the specific functions 
\bes\ben
W&=&\phi-\phi^3+\frac35\phi^5-\frac17\phi^7\label{k1}
\\
W&=&\phi\sqrt{1-\phi^2}\left(\frac58-\frac14\phi^2\right)+\frac38 {\rm arcsin}(\phi)\label{c1}
\een\ees
which give the following potentials
\bes\ben
V(\phi)&=&\frac{3}{4}(1-\phi^2)^4 \label{ppe1}
\\
V(\phi)&=&\frac{3}{4}(1-\phi^2)^2\label{ppe2}
\een\ees
respectively, which we plot in Fig.~[2].

\begin{figure}[ht!]
\centering
\includegraphics[width=5cm]{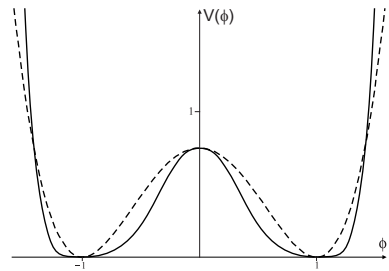}
\centering
\caption{Plots of the potentials of the models given by \eqref{ppe1} (solid line) and \eqref{ppe2} (dashed line).}
\end{figure}

In the first case the model will lead us with kinklike solution. In the second case the field behaves differently, and the model will lead us with compactons, in a way similar to the case considered in \cite{adam}.

Let us now study the model described by \eqref{k1}. The first-order equation is $\phi^\prime=1-\phi^2$ and we get the kinklike solution
$\phi(x)=\tanh(x)$, with energy $E=\Delta W=32/35$. For the fluctuations, the potential in the Schr\"odinger-like equation gets the form
\ben\label{pp1}
U(z)=48-60\,{\rm sech}^2(\sqrt{3}z)
\een
This is the modified P\"oschl-Teller potential \cite{PT} and the normalized zero mode is now
\ben\label{zero1}
u_0(z)=\frac18\sqrt{70\sqrt{3}}\;{\rm sech}^4(\sqrt{3}z)
\een
In Fig.~3 we plot both $U(z)$ and the zero mode $u_0(z)$. This potential supports the zero mode and other three bound states, with energies $E_r=3r(8-r),$ for $r=0,1,2,3.$

\begin{figure}[ht!]
\centering
\includegraphics[width=5cm]{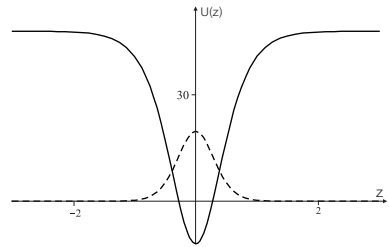}
\centering
\caption{Plots of the potential (solid line) and zero mode (dashed line) given by \eqref{pp1} and \eqref{zero1}, respectively. }
\end{figure}

The other model is described by \eqref{c1}. The first-order equation is $\phi^\prime=(1-\phi^2)^{1/2},$ and may support compactons \cite{CCC}, which
are growing in importance in a diversity of scenarios, in particular from the point of view of pattern formation, since patterns usually appear in nature with finite extent. In the present work, the explicit solution of the first-order equation is
\ben
\phi(x)=\left\{\begin{array}{lll}0, \;\;\;{\rm for}\;\;\;x<-\frac{\pi}{2}
\\\sin(x),\;\;\; {\rm for}\;\;\; -\frac{\pi}{2}\leq x\leq\frac{\pi}{2}
\\0,\;\;\; {\rm for}\;\;\; x>\frac{\pi}{2}\end{array}\right. 
\een
The corresponding energy is $E=3\pi/8.$ In this case the potential and the normalized zero mode are given by
\ben\label{pp2}
U(z)\!=\!\left\{\begin{array}{lll}\infty,\;\;\;{\rm for}\;\;\;z<-\frac{\pi}{2\sqrt{3}}
\\-12+6\sec^2(\sqrt{3}z),\;{\rm for}\;-\frac{\pi}{2\sqrt{3}}\leq z\leq\frac{\pi}{2\sqrt{3}}
\\ \infty,\;\;\;{\rm for}\;\;\;z>\frac{\pi}{2\sqrt{3}}\end{array}\right.
\een
and
\be\label{zero2}
u_0(z)\!=\!\sqrt{\frac{2^3}{\pi\sqrt{3}}}\left\{\begin{array}{lll}0,\;\;\;{\rm for}\;\;\;z<-\frac{\pi}{2\sqrt{3}}
\\ \cos^2(\sqrt{3}z),\;{\rm for}\;-\frac{\pi}{2\sqrt{3}}\leq z\leq\frac{\pi}{2\sqrt{3}}
\\0,\;\;\;{\rm for}\;\;\;z>\frac{\pi}{2\sqrt{3}}\end{array}\right.
\ee
The potential in \eqref{pp2} is the P\"oschl-Teller potential \cite{PT}, which is plotted in Fig.~4, together with the above zero mode.

\begin{figure}[ht!]
\centering
\includegraphics[width=5cm]{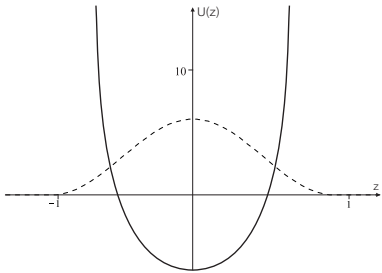}
\centering
\caption{Plots of the potential (solid line) and zero mode (dashed line) given by \eqref{pp2} and \eqref{zero2}, respectively.}
\end{figure}

The interesting feature of the potential in \eqref{pp2} is that it only supports bound states, and for $r=0,1,2,...,$ the corresponding eigenvalues are given by $E_r=12r(r+2).$ Its behavior is similar to the potentials found in \cite{blmo} for models of the form ${\cal L}=V(\phi)F(X),$ which appears in the case of tachyons and their generalizations.

The case of compactons may lead to an hybrid configuration, named half-compacton, as noted in \cite{adam} for the model with the potential
\be\label{CC1p}
V(\phi)=\frac34(1-\phi)^2(1+\phi)^4
\ee 
which is plotted in Fig.~5. The solution has the form
\ben
\phi(x)=\left\{\begin{array}{ll}1-2\,\tanh^2\left(\frac{x\sqrt{2}}{2}\right),\;{\rm for}\;x>0
\\1,\;\;{\rm for}\;\;x\leq0\end{array}\right.
\een
Here we get
\be
W=-\frac{2}{1155}(533+755\phi-455\phi^2+105\phi^3)(1-\phi)^{5/2}
\ee
The potential for the Schr\"odinger-like equation is such that
\ben\label{CC1}
U(z)=\left\{\begin{array}{ll}3\frac{8-28\,{\rm sech}^2\left(\frac{\sqrt{6}}{2}z\right)+21\,{\rm sech}^4\left(\frac{\sqrt{6}}{2}z\right)}{\tanh^2\left(\frac{\sqrt{6}}{2}z\right)},\;{\rm for}\; z>0\\
\infty,\; {\rm for} \;z\leq0\end{array}\right.
\een
In this case, $U(z)$ has an hybrid asymptotic behavior, having P\"oschl-Teller and modified P\"oschl-Teller tails, as we show in Fig.~6.
The energy of the hybrid configuration is $E=1024\sqrt{2}/1155,$ and the zero mode has the form  
\be\label{zeroCC1}
u_0(z)=\left\{\begin{array}{ll}\sqrt{\frac{5\cdot7\cdot11}{\sqrt{3\cdot2^{21}}}}\tanh^2\left(\frac{\sqrt{6}}{2}z\right)\,{\rm sech}^4\left(\frac{\sqrt{6}}{2}z\right),\;{\rm for}\;z>0
\\ 0,\;{\rm for}\;z\leq0 \end{array}\right.
\ee
which is also plotted in Fig.~6. In this case, the reflection coefficient is unit, and this would modify the standard behavior concerning the scattering of particles by the wall itself, changing the frictional force which would act on the wall, introducing new features in the wall evolution in a $(3,1)$ cosmological scenario \cite{books}, but this is out of the scope of this work. 

\begin{figure}[ht!]
\centering
\includegraphics[width=5cm]{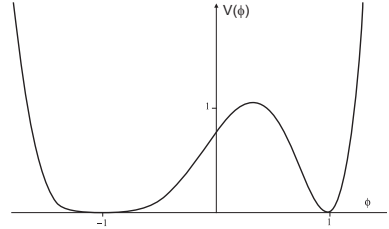}
\centering
\caption{Plot of the potential given by \eqref{CC1p}.}
\end{figure}

\begin{figure}[ht!]
\centering
\includegraphics[width=5cm]{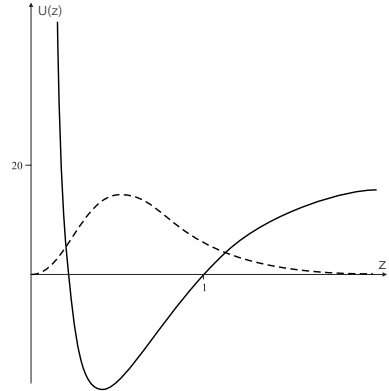}
\centering
\caption{Plots of the potential (solid line) and zero mode (dashed line) given by \eqref{CC1} and \eqref{zeroCC1}, respectively. }
\end{figure}
{\bf IV. Ending comments.} In this work we have shown how to extend the first-order formalism to generalized models, described by scalar fields with non standard dynamics. The main issue here is the introduction of $W$ and the related first-order equations of motion \eqref{W}. These equations together with the stressless condition \eqref{condstress} simplify the investigations of linearly stable defect solutions, which in specific conditions can be obtained analytically.

We have investigated several models, in general described by a set of real scalar fields. In the case of a single field, we have dealt with the Lagrange density of the form $F(X)-V(\phi)$, and we have considered the cases where $F(X)$ is governed by $X+\alpha X|X|$ or $X|X|^{n-1}$. We have also considered several explicit forms for $W$, in particular the case with $W=\phi-\phi^3/3,$ which reproduces the standard $\phi^4$ model if the kinetic term is also of the standard form $F(X)=X.$ The general result is that topological defects also appear in models with generalized dynamics, and there it is also possible to formulate a first-order framework in which one deals with first-order differential equations, which solve the corresponding equations of motion and ease the study concerning the search for analytical results. 

In the present work, we have focused mainly on explicit models governed by a single real scalar field. Evidently, we can also deal with two or more fields, and we will return to the subject in a future work, where we investigate more sophisticated models, governed by two or more real or complex scalar fields. In particular, we will deal with generalized models constructed from complex fields $\varphi_i,$ in which the complex fields have the form $\varphi_i=\phi_i+i\,\chi_i$. In this case, the function $F(X)$ should deal with $X=\partial_\mu{\ov\varphi_i}\partial^\mu{\varphi_i},$ with the sum over
all the complex fields implied. This form is specific but important, since it allows generalizing the dynamics one usually finds in the standard Wess-Zumino model, which has interesting features which will certainly help us to make explicit integrations \cite{wz}. 

The presence of topological defects in models described by scalar fields is well known in high energy physics \cite{books}, and the present work contributes to show how they appear in models where the dynamics is generalized to include higher order contributions on $X.$ The results are of current interest to high energy physics, since nowadays there are situations in which modifications of the standard dynamics are welcome.


\end{document}